\documentclass[aps,prl,singlecolumn,superscriptaddress,notitlepage]{revtex4-2}

\usepackage{amsmath}
\usepackage{graphicx}

\begin{document}

\title[In-plane resonant excitation of QDs in a dual-mode PCW]{In-plane resonant excitation of quantum dots in a dual-mode photonic-crystal waveguide with high $\beta$-factor}

\author{Xiaoyan Zhou}
\email{xiaoyan\_zhou@tju.edu.cn}
\affiliation{Key Laboratory of Opto-Electronics Information Technology of Ministry of Education, School of Precision Instruments and Opto-Electronics Engineering, Tianjin University, 300072 Tianjin, China}
\affiliation{Center for Hybrid Quantum Networks (Hy-Q), Niels Bohr Institute, University of Copenhagen, Blegdamsvej 17, DK-2100 Copenhagen, Denmark}
\author{Peter Lodahl}
\affiliation{Center for Hybrid Quantum Networks (Hy-Q), Niels Bohr Institute, University of Copenhagen, Blegdamsvej 17, DK-2100 Copenhagen, Denmark}
\author{Leonardo Midolo}
\email{midolo@nbi.ku.dk}
\affiliation{Center for Hybrid Quantum Networks (Hy-Q), Niels Bohr Institute, University of Copenhagen, Blegdamsvej 17, DK-2100 Copenhagen, Denmark}

\begin{abstract}
A high-quality quantum dot (QD) single-photon source is a key resource for quantum information processing. Exciting a QD emitter resonantly can greatly suppress decoherence processes and lead to highly indistinguishable single-photon generation. It has, however, remained a challenge to implement strict resonant excitation in a stable and scalable way, without compromising any of the key specs of the source (efficiency, purity, and indistinguishability). In this work, we propose a novel dual-mode photonic-crystal waveguide that realizes direct in-plane resonant excitation of the embedded QDs. The device relies on a two-mode waveguide design, which allows exploiting one mode for excitation of the QD and the other mode for collecting the emitted single photons with high efficiency. By proper engineering of the photonic bandstructure, we propose a design with single-photon collection efficiency of $\beta > 0.95$ together with a single-photon impurity of $\epsilon< 5 \times 10^{-3}$ over a broad spectral and spatial range. The device has a compact footprint of $\sim 50$ $\mu$m$^2$ and would enable stable and scalable excitation of multiple emitters for multi-photon quantum applications.
\end{abstract}

\maketitle

\section{Introduction}\label{sec:intro}
Quantum information technologies pose stringent requirements for a single-photon source, which should have the simultaneous merits of being deterministic, pure, indistinguishable, and bright \cite{uppu2021nn}. Among others, quantum dots (QDs) have emerged as nearly ideal quantum emitters meeting all the above-mentioned requirements \cite{senellart2017natnano,uppu2020scienceadv,tomm2021nn}. Indistinguishable single-photon generation has been achieved with the combination of electric contacts and the resonant excitation scheme, which greatly suppresses the decoherence and charge noise \cite{uppu2020scienceadv,tomm2021nn}. Embedding QDs in specially designed nanostructures enables a significantly enhanced collection efficiency into a single optical mode and noise suppression through the Purcell effect \cite{Ding2016prl,pedersen2020ACSphoton,uppu2020scienceadv,tomm2021nn}. In particular, a planar photonic-crystal waveguide (PCW) fabricated around the QD offers the additional advantages of a broadband enhancement of collection efficiency to almost unity ($\beta > 0.98$) \cite{Arcari2014prl,Javadi2018josab,Scarpelli2019prb}, and provides a natural platform for integration with other functional photonic modules, such as photonic switches, filters, nonlinear units etc., which enable the scaling-up of QD single-photon sources to more advanced quantum functionalities \cite{Hepp2019aqt,Elshaari2020np,uppu2021nn}.

One of the main obstacles in performing resonant excitation schemes with QDs lies in separating the excitation laser from the collected photons without compromising the source performance. A common approach, employed in, e.g., micro-pillar cavities, is to cross-polarize the excitation laser with respect to the collection of the emitted single photons in a confocal configuration. Unfortunately, this method leads to an intrinsic loss of efficiency although the use of elliptical cavities has been implemented to alleviate this deficit \cite{Wang2019natphton,Wang2020prb}. In contrast, spatial extinction can be implemented in planar QD devices where the QDs are excited via leaky modes and the emitted photons are collected into the waveguide mode \cite{Monniello2014prb,Arcari2014prl,Javadi2018josab,Ding2019prapp}. However, it is challenging to realize stable excitation in this way, as the radiation modes have a position-dependent distribution and polarization due to the nanostructured surfaces \cite{Ding2019prapp}. These deficits may limit the scalability of the technology, since it will be challenging to realize multiple QD experiments.

To improve the stability and scalability it is therefore desirable to implement a scheme for resonant excitation that uses exclusively modes propagating in the waveguide for excitation and collection. Recently, a novel in-plane excitation scheme for a QD in a multi-mode nanobeam waveguide (NW) was realized by exciting the emitters via the first-order waveguide mode and collecting the emitted photons in the fundamental mode \cite{Uppu2020nc}. Here high purity and photon indistinguishability was successfully achieved together with the ``hands-free'' operation of the device over 110 hours. However, in these devices the internal single-photon efficiency ($\beta \sim 0.80$) was intrinsically limited by the NW geometry. Therefore, building a deterministic source with near-unity efficiency, for applications such as device-independent quantum key distribution \cite{kolodynski2020quantum,gonzalezruiz2021arx}, requires developing a new scheme for in-plane resonant excitation.

In this work, we propose a novel dual-mode PCW to realize simultaneously high extinction of the excitation laser and efficient collection of the emitted single photons. The device is schematically shown in Fig. 1(a). A mode-filter PCW and a W1 PCW (i.e. a standard single-line defect PCW) are connected to the two sides of a dual-mode PCW, where one mode is employed for laser excitation and the other for photon collection. We carry out numerical simulations of the device and show that near-unity $\beta$-factor ($\beta \sim 0.98$) and low single-photon impurity $\epsilon \sim 5\times10^{-3}$ can be simultaneously achieved for optimal QD locations. Moreover, high $\beta > 0.95$ and single-photon purity of $\epsilon < 5\times10^{-3}$ can be achieved spectrally over a broad wavelength range of 10 nm, and spatially for more than 22\% of the effective area of the waveguide. Besides, a mode adapter to the output waveguide has been optimized to obtain an overall collection efficiency of 0.89. The proposed device is compact (footprint is 13.2 $\mu m \times$ 4 $\mu$m) and the excitation scheme is potentially very stable, opening new possibilities for scalable quantum technologies \cite{Wang2019prl,uppu2021nn}.

\begin{figure}
    \centering
    \includegraphics[width=13cm]{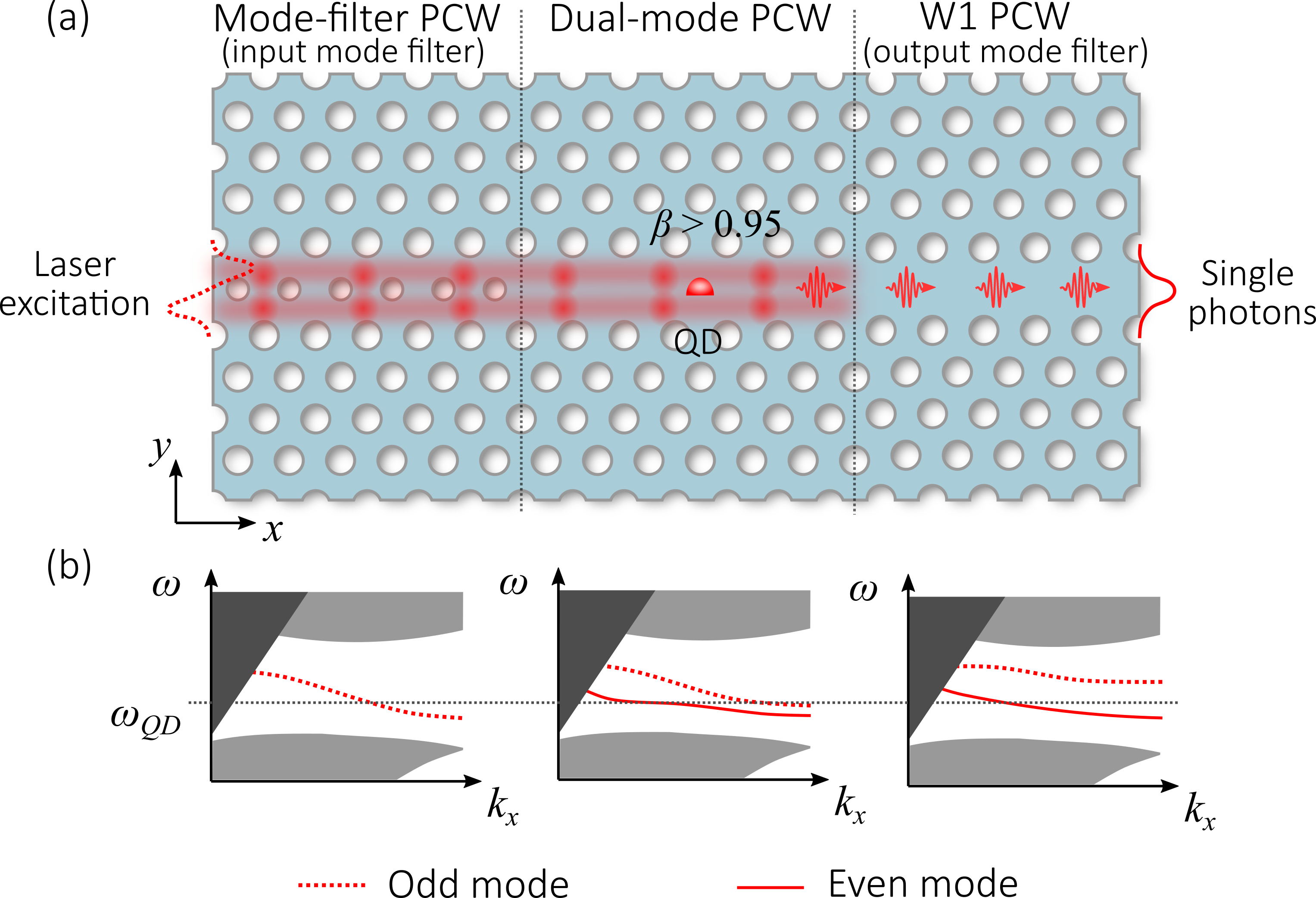}
    \caption{In-plane resonant excitation scheme of a quantum dot (QD) in a dual-mode photonic-crystal waveguide (PCW). (a) After passing through the input mode-filter PCW, the pump laser at $\omega_{QD}$ can only propagate in the odd mode of the dual-mode PCW. The excited QD emits single photons at the same frequency ($\omega_{QD}$), and is coupled to the even mode of the dual-mode PCW with near-unity efficiency, i.e., $\beta > 0.95$, while the laser in the odd mode is reflected at the interface to the third section of the device that consists of a regular W1 PCW. Odd and even modes are named according to their symmetry in the $y$-direction. (b) Photonic band structures for the three sections. The light grey and the dark grey regions show the continuous photonic-crystal modes and the radiation modes, respectively. At the working frequency $\omega_{QD}$, the input mode-filter PCW supports exclusively an odd waveguide mode, while the dual-mode PCW supports both an even mode and an odd mode. The output W1 PCW is designed to allow the even mode only and extinguish the pump laser in the odd mode.}
    \label{fig:schematic}
\end{figure}

\section{Design principle and band-structure engineering}\label{sec:bands}
Before presenting the detailed design parameters, we briefly introduce the design principle of the excitation scheme. The PCW can be divided into three sections, namely the mode-filter PCW, the dual-mode PCW, and the W1 PCW, as shown from left to right in Fig.~\ref{fig:schematic}(a). The key section is the dual-mode PCW, where the emitter is located. It is designed to support two guided modes with even and odd symmetry along the $y$-direction at the QD emission frequency $\omega_{QD}$ (cf. middle panel of Fig.~\ref{fig:schematic}(b)). Under resonant excitation, the pump laser excites the QD via the (weakly coupled) odd mode of the PCW, while the emitted photons couple mainly to the even mode. To pump the dual-mode PCW with the odd mode, a mode-filter PCW, featuring an additional row of holes in the middle of the waveguide, has been designed. The right side of the dual-mode PCW is connected to another mode filter, i.e., the W1 PCW, which is single-mode at $\omega_{QD}$ and only transmits the even mode to which most of the emitted single photons are coupled. The photonic band structures of the various sections are schematically shown in  Fig.~\ref{fig:schematic}(b). Such an in-plane excitation scheme provides similar functionality reported in Ref.~\cite{Uppu2020nc}, but allows for a much higher $\beta$-factor and a much more compact device footprint.

To achieve high $\beta$-factor, it is desirable that the emitted photons couple primarily to a single mode, which requires careful band-structure engineering of the dual-mode PCW. It has been previously shown that near-unity $\beta$-factor can be achieved for QDs in a single-mode PCW working close to the photonic band edge\cite{Arcari2014prl}, where the emission rate of the QD is Purcell-enhanced due to the large density of states (DOS) and the coupling to radiation modes is strongly suppressed by virtue of the photonic band gap \cite{lodahl2015rmp}. Although the dual-mode PCW supports two modes, it is possible to retain the high $\beta$-factor to the even mode (the collection mode) by engineering the dispersion of the two modes, such that the DOS of the even mode at the emitter frequency $\omega_{QD}$ is significantly higher than the odd mode (i.e. the excitation mode), as shown in the middle panel of Fig.~\ref{fig:schematic}(b).

\begin{figure}
    \centering
    \includegraphics[width=13cm]{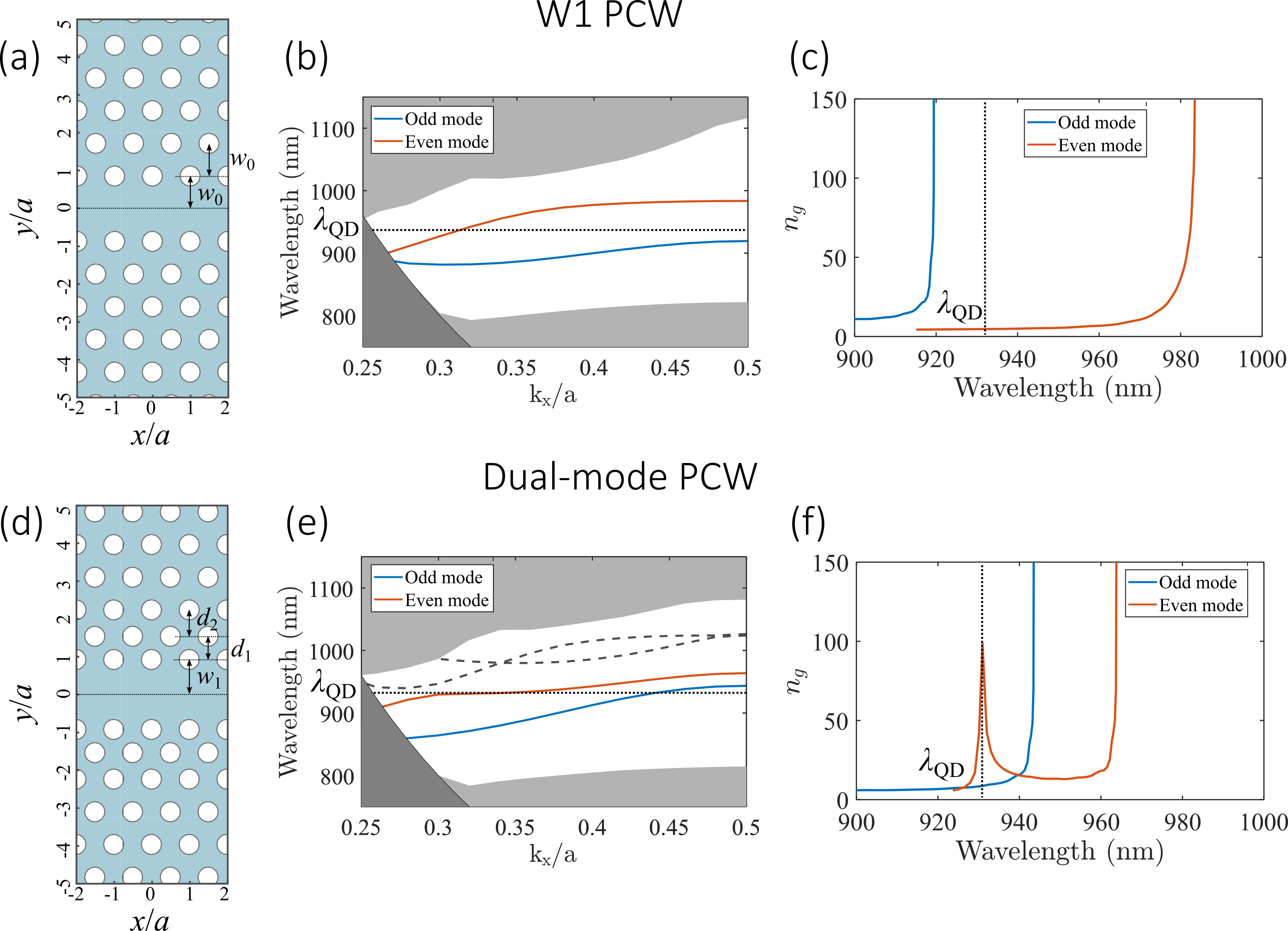}
    \caption{Band-structure engineering of the dual-mode PCW compared to a W1 PCW. (a) Geometry of a W1 PCW ($w_0 = a\sqrt{3}/2$), and (b) its band structure. (c) The group index ($n_g$) of the two waveguide bands of the W1 PCW as a function of wavelength. (d) Geometry of the designed dual-mode PCW, with modified $y$-coordinates of the three rows of holes closest to the waveguide ($w_1 = 1.07 w_0$, $d_1 = w_0 - 60$ nm, and $d_2 = w_0 - 40$ nm), and (e) its band structure. The dashed lines represent other guided modes. (f) $n_g$ of the two waveguide modes of the dual-mode PCW as a function of wavelength, showing a peak in the even mode at the QD wavelength ($\lambda_{QD}$). The grey and dark grey areas in (b) and (e) represent the continuous photonic-crystal modes and the radiation modes, respectively.
    }
    \label{fig:band_structure}
\end{figure}

We designed the dual-mode PCW starting from a standard W1 PCW, as shown in Fig.~\ref{fig:band_structure}(a). The W1 PCW is a single-line defect waveguide formed in a quasi-2D photonic-crystal membrane with a triangular lattice of circular holes. The distance between the center of the waveguide and the row of holes adjacent to the waveguide is $w_0 = a\sqrt{3}/2$. The lattice constant, $a =  240$ nm, and the radii of the holes, $r_0 = 64$ nm, are chosen for single-mode operation around the typical QD emission wavelength ($\lambda_{QD}=930$ nm) \cite{lodahl2015rmp}. In the numerical simulations, the membrane thickness is set to $t = 175$ nm for the standard \textit{p-i-n} doped samples used in previous works \cite{Uppu2020nc}, and the refractive index of GaAs at cryogenic temperature $n_{GaAs} = 3.475$ is used.

The band structure of the W1 PCW is calculated using a 3D finite-element method (FEM) solver and is shown in Fig.~\ref{fig:band_structure}(b). There are two waveguide modes in the photonic band gap, which can be classified depending on their symmetry along the $y$-direction as even fundamental mode and odd first-order mode. We note that in the W1 PCW only the even mode is allowed at $\lambda_{QD}$. The group-index $n_g = c/v_g$, where $c$ is the speed of light in vacuum and $v_g$ is the group velocity of the photonic mode, is proportional to the DOS determining the strength of light-matter interaction \cite{lodahl2015rmp}. The calculated group indices of the two modes are shown in Fig.~\ref{fig:band_structure}(c), and they feature a rapid increase near the PCW band edges, i.e., at $k_x/a \sim 0.5$.

To engineer the waveguide bands and create the desired dual-mode PCW, we modify the positions of the holes around the waveguide. We increase the distance between the inner rows of holes, i.e., $w_1 > w_0$ (Fig.~\ref{fig:band_structure}(d)), to allow both odd and even modes to propagate at $\lambda_{QD}$. To increase the DOS for the even mode, the PCW band needs to be engineered such that a high group index ($n_{g1}$) is achieved at smaller in-plane wave-number $k_x$, away from the edge of the first Brillouin zone. We create a non-monotonic group index spectrum by modifying the distance $d_1$ between the 1st and the 2nd rows of holes, and the distance $d_2$ between the 2nd and the 3rd row of holes (cf. Fig. \ref{fig:band_structure}(d)). Similar techniques have been used in previous works to expand the slow-light bandwidth in single-mode PCWs \cite{Li08oe}. An optimized $n_{g1} = 98.5$ at $\lambda_{QD} = 931$ nm and a broad dual-mode spectral band are achieved for the designed structure. Importantly, the group index of the even mode is 11 times larger than the odd mode ($n_{g2} = 8.7$) at $\lambda_{QD}$. In practice, when working in such a high $n_g$ region, the emitted photons might suffer from considerable back-scattering from fabrication imperfections, resulting in the so-called Anderson localization effect \cite{lodahl2015rmp,Arcari2014prl}. This effect can be largely avoided by designing the dual-mode PCW to be shorter than the localization length, typically $< 10$ $\mu $m \cite{lodahl2015rmp}.

\begin{figure}
    \centering
    \includegraphics[width=13cm]{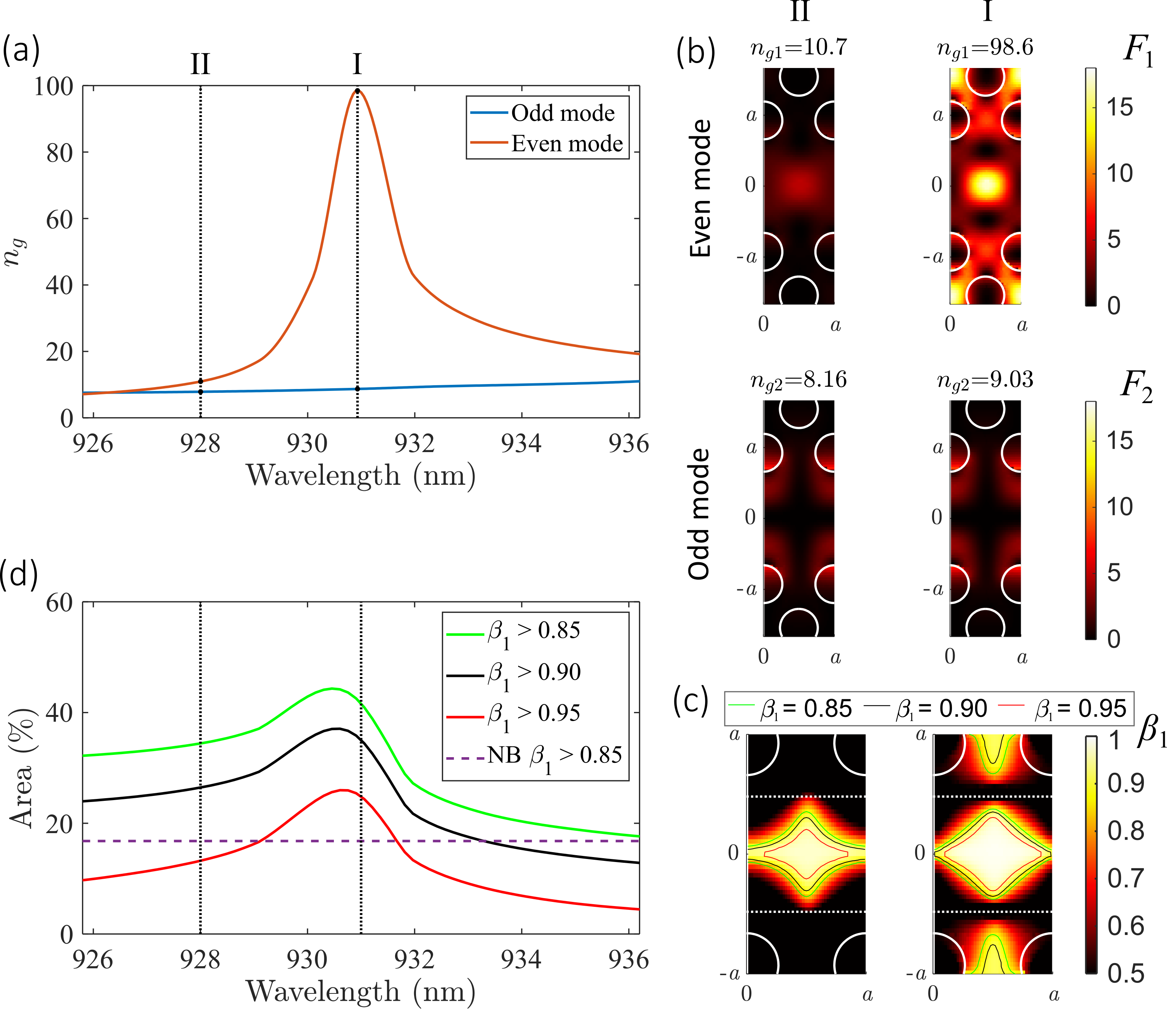}
    \caption{(a) Group index of the two modes of the dual-mode PCW as a function of wavelength. Spatial maps of (b) Purcell factor for the even mode ($F_{1}$) and the odd mode ($F_{2}$) and (c) $\beta$-factor for the even mode ($\beta_1$) at 931 nm and 928 nm labelled as I and II in (a). (d) Fraction of the PCW unit cell where $\beta_1$ is larger than 0.85 (blue line), 0.90 (black line), and 0.95 (red line) as a function of wavelength, compared with NW with $\beta_1>0.85$ (dashed line). }

    \label{fig:maps}
\end{figure}

To estimate the $\beta$-factor of the emitter, the local density of states (LDOS) or the position-dependent Purcell factor of the mode, is calculated.
The Purcell factor $F_{j}(\mathbf{r_d})$ to a specific mode $j$ at the emitter dipole location ($\mathbf{r_d}$), is given by \cite{Hughes04ol}:

\begin{equation}
    F_{j}(\mathbf{r_d}) = \frac{3 \pi c^2 a n_{gj}|\mathbf{E_j(r_d)} \cdot \mathbf{n_y}|^2}{n_{GaAs}\omega^2},
\end{equation}

where $j=1, 2$ refers to the even and odd modes, respectively, $\mathbf{E_j(r_d)}$ is the normalized electric field at $\mathbf{r_d}$ integrated over the unit cell, i.e.,  $\int_{unit-cell} |\mathbf{E_j(r)}|^2 \epsilon_r(\mathbf{r})d\mathbf{r} = 1$, where $\epsilon_r(\mathbf{r})$ is the relative electric permittivity, and $\mathbf{n_y}$ is the unit vector parallel to the $y$-polarized dipole emitter. The Purcell factor is proportional to both the group index and the intensity of the electric field component parallel to the transition dipole. The $\beta$-factor of the emitter coupling to a specific waveguide mode in the dual-mode PCW is related to the Purcell factor by \cite{MangaRao2007prb}

\begin{equation}
    \beta_j(\mathbf{r_d}) = \frac{F_{j}(\mathbf{r_d})}{F_{1}(\mathbf{r_d}) + F_{2}(\mathbf{r_d}) + F_{ng}(\mathbf{r_d})},
\end{equation}

where $F_{ng}(\mathbf{r_d})$ is the Purcell factor of the emitter coupling to the non-guided (or leaky) modes. As reported in Ref. \cite{Javadi2018josab}, the PCW greatly suppresses the coupling to the non-guided modes, resulting in $F_{ng}<<1$. Consequently, in cases where $F_{1} >> F_{2}$ and $F_{1}>>1$, $\beta_1$ reaches almost unity, implying very high collection efficiency to the even PCW mode.

We calculate the Purcell factors with a FEM eigenfrequency solver at $\lambda = 931$ nm with largest $n_{g1} = 98.6$, and, for comparison, at $\lambda = 928$ nm with $n_{g1} = 10.7$, labelled as I and II in Fig.~\ref{fig:maps}(a). The resulting Purcell distribution maps are shown in Fig.~\ref{fig:maps}(b). From these maps, it is evident that, in the center of the PCW, the condition $F_{1}>>1>>F_{2}$ is fulfilled. As a consequence, large $\beta_{1}$ can be obtained for nearly-centered QDs, as shown in Fig.~\ref{fig:maps}(c). Here, the $\beta$-factor was calculated assuming $F_{ng} = 0.13$ in Eq.~(2), which sets the upper-bound of the emission rate to the radiation modes in a PCW \cite{Javadi2018josab}. $\beta_1$ exceeds 0.95 for a broad wavelength range and for emitters positioned within a large fraction of the unit cell. These values are comparable with what is typically achieved in standard single-mode PCWs \cite{Arcari2014prl,Javadi2018josab,Scarpelli2019prb} but much larger than the best theoretically attainable value in dual-mode NWs (i.e. around 0.90) \cite{Uppu2020nc}.

To estimate the probability of finding an emitter located in a high-$\beta$ position within the waveguide, we calculate the area ($A_{\beta}$) where $\beta$ is higher than a certain threshold. The area is normalized to an effective unit cell area $A_{eff}$, defined by the white dotted lines in Fig.~\ref{fig:maps}(c). The QDs located within this area are sufficiently far (>43 nm) from the waveguide holes, i.e., from etched surfaces, and are thus expected to exhibit near-transform limited behavior \cite{pedersen2020ACSphoton}. At the optimal wavelength, there is a 26\% chance of finding an emitter with very high $\beta_1$ of $>0.95$ in the PCW, which is not achievable for the dual-mode NW \cite{Uppu2020nc}. The chance improves further to 37\% for $\beta_1>0.90$ and 44\% for $\beta_1>0.85$ at the optimal wavelength, which is more than twice the value in dual-mode NWs. Moreover, the device shows broadband operation capability, and the likelihood of finding a QD with $\beta_1>0.85$ is larger than the NW over the 10-nm wavelength range, making the design robust to fabrication imperfections.

\section{Mode filters and adapters for efficient pump laser suppression and single-photon collection}\label{sec:trans}

\begin{figure}
    \centering
    \includegraphics[width=13cm]{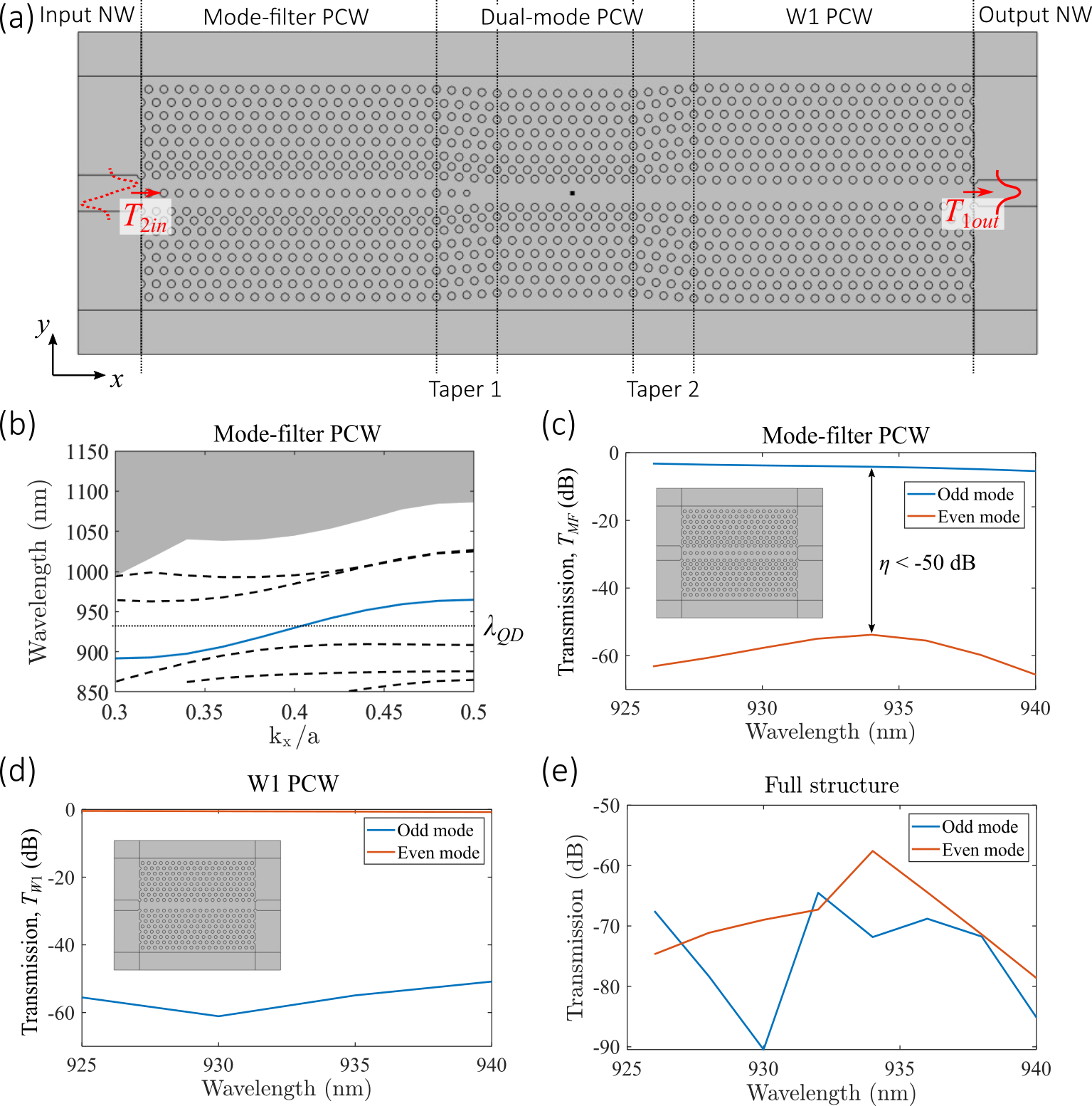}
    \caption{
     Device design for achieving high extinction of the pump laser and high collection efficiency of the emitted single photons. (a) Schematic of the full structure: the functional PCW sections consist of the input mode-filter PCW, the dual-mode PCW, the output W1 PCW, and two short tapers. The PCW structure is connected by an input NW and an output NW. $T_{2in}$ and $T_{1out}$ represent the transmission for the odd mode and even mode at the input and output PCW-NW interfaces, respectively. (b) The band structure of the mode-filter PCW, showing that only the odd mode is supported around the operation wavelength $\lambda_{QD}$. The transmission of the odd and even modes for (c) the isolated mode-filter PCW and (d) the isolated W1 PCW. The insets show the structures used in simulation. (e) The transmission of the full structure in (a) shows a large extinction ($< 5 \times 10^{-5}$) for both the odd or even input modes.}
    \label{fig:full_structure}
\end{figure}

Ensuring a high extinction of the excitation laser is of great importance to suppress the multi-photon probability of the single-photon source. The single-photon impurity is defined as $\epsilon = I_{res}/I_{ph}$, which describes the number of residual laser photons in the collection ($I_{res}$) compared to the emitted single photon ($I_{ph}$). Large laser extinction can be achieved by adding two mode filters before and after the dual-mode section. While pump laser in the odd mode can be easily extinguished with a standard W1 PCW (or single-mode NW) that only supports the even mode, filtering out even-mode excitation laser is typically much more challenging. Assuming emitted single photons are eventually collected into a single-mode fiber, which is typically the case, any laser residual or single photons in the odd mode at the output of the PCW structure can be considered fully suppressed. Thus, $I_{res}$ and $I_{ph}$ can be estimated as $I_{res} \approx I_{l1}T_{1in}T_{1out}$ and $I_{ph} \approx I_{l2}T_{2in} \beta_2\beta_1 T_{1out}$ in the limit of weak excitation, where $I_{l1}$ and $I_{l2}$ represent pump laser in the even and odd modes of the input NW, $T_{1in}$ and $T_{2in}$ are the transmission of the even and odd modes for the input mode-filter PCW, and $T_{1out}$ is the even-mode transmission for the output W1 PCW. We define $\eta=I_{l1}T_{1in}/(I_{l2}T_{2in})$, then $\epsilon$ can be expressed as $\epsilon = I_{res}/I_{ph} = I_{l1}T_{1in}/(I_{l2}T_{2in}\beta_2\beta_1) = \eta/(\beta_2 \beta_1)$. One notes immediately that $\eta$ is only limited by the ratio of laser in the even mode compared to the odd mode in the dual-mode PCW section. Minimizing $\epsilon$ while keeping $\beta_1$ close to unity requires achieving high extinction $\eta$, ideally around $10^{-4}$--$10^{-6}$.

The mode filters are shown in Fig.~\ref{fig:full_structure} (a), connected to the dual-mode waveguide via tapered mode adaptors that adiabatically morph the unit cells from one type into the next.
Here we assume that the excitation laser is launched into both the even and odd modes at the input of the structure with a $\sim$ 50\% ratio ($I_{1l} = I_{l2}$), for example by exciting a Y-junction from one of its branches as in Ref. \cite{Uppu2020nc}.
We design the even-mode filter PCW by widening the waveguide of the dual-mode PCW further so that $w_1 = 1.38w_0$, and add a row of holes with the same radii of $r_0$ in the center of the waveguide. We confirm that such a mode filter only support an odd PCW mode around the operation wavelength $\lambda_{QD}$, by calculating the photonic bands plotted in Fig.~\ref{fig:full_structure}(b). The transmission of the even and odd mode are calculated by interfacing the mode-filter PCW with two NWs (inset of Fig.~\ref{fig:full_structure}(c)), and by using either the odd or the even NW mode as input. Figure~\ref{fig:full_structure}(c) shows that the transmission of the even mode can be suppressed to $T_{1in} < 4\times10^{-6}$ (i.e. an extinction of -54 dB) over a $15$-nm bandwidth around $\lambda_{QD}$, while the transmission efficiency of the odd mode at $\lambda_{QD}$ is $T_{MF-odd} = 0.4$. We note that the loss in the odd mode mainly stems from the mode mismatch at the two interfaces between the NW and the PCW, thus we estimate $T_{2in} \approx \sqrt{T_{MF-odd}} = 0.63$ for a single interface. The extinction given by the mode-filter PCW section is thus $\eta \approx T_{1in}/T_{2in} < -50$ dB.

In a similar way, we simulate the transmission of the odd-mode filter. This is a standard W1 PCW designed so that $\lambda_{QD}$ sits in the ``fast-light'' region of the even mode with a low $n_g\sim 5$ to avoid losses at the PCW-NW interface, which is crucial for achieving high collection efficiency of the emitted single photons \cite{Arcari2014prl,Javadi2015nc}. Figure~\ref{fig:full_structure}(d) shows that the transmission of the even mode for the structure in the inset $T_{W1-even}$ is around 0.90 at $\lambda_{QD}$, implying the single-interface transmission would reach $T_{1out} \approx \sqrt{T_{W1-even}} = 0.95$. The transmission for the odd mode $T_{2out}$ is suppressed to $10^{-6}$ over the full wavelength range.

The overall laser extinction is calculated with a 3D FEM frequency-domain transmission simulation. Figure~\ref{fig:full_structure}(e) shows that the overall structure exhibits simultaneous suppression of the pump laser in both modes, where the even mode is rejected by the mode-filter PCW and the odd mode is rejected by the W1 PCW after exciting the QD emitter.

\begin{figure}
    \centering
    \includegraphics[width=13cm]{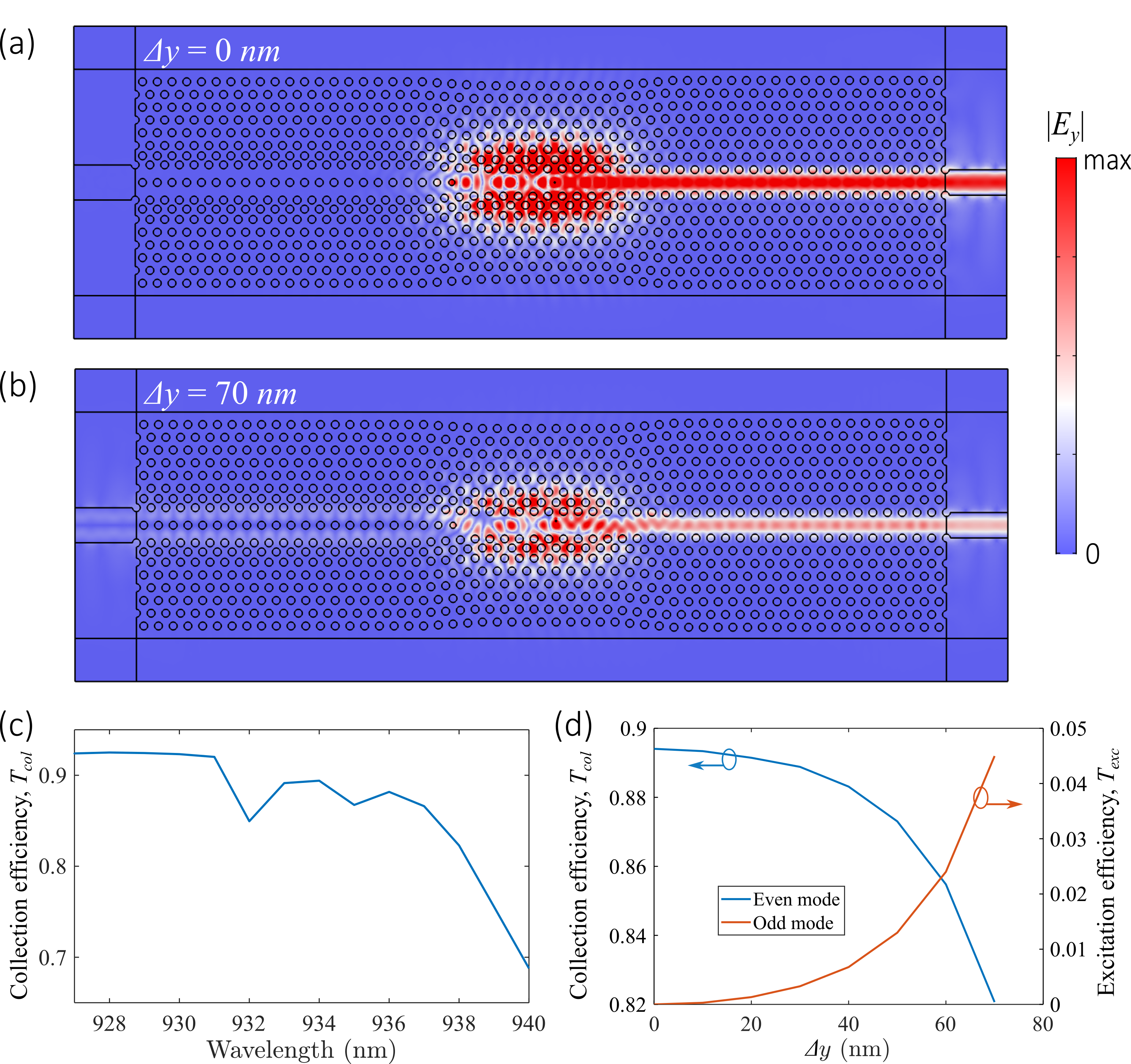}
    \caption{High overall collection efficiency at different wavelengths and a broad area. (a) and (b) Field amplitude of $E_y$ generated by a dipole emitter ($\lambda_{QD} = 934$ nm) in dual-mode PCW at the center ($\Delta y = 0$ nm) and with an offset in $y$ ($\Delta y = 70$ nm), respectively. (c) Collection efficiency defined as $T_{col} = P_{out}/P_{tot}$ as a function of wavelength for the centered dipole emitter ($\Delta y = 0$), where $P_{out}$ and $P_{tot}$ are the power transmitted to the even mode in the output NW and total power emitted by the dipole, respectively. (d) Collection efficiency decreases and the excitation efficiency, defined as the $T_{exc} = P_{in}/P_{tot}$ where $P_{in}$ is the power transmitted to the odd mode in the input NW, increases with the offset of the dipole ($\Delta y$).}
    \label{fig:dipole}
\end{figure}

To optimize the photon collection efficiency, we add two taper sections (labeled Taper 1 and Taper 2 in Fig.~\ref{fig:full_structure}(a)) and adjust the geometric parameters to reduce any scattering loss caused by mode mismatch at the PCW interfaces. The optimization is done by simulating a dipole emitter located at the center of the dual-mode PCW as a source, and by monitoring the collection efficiency defined as $T_{col} = P_{out}/P_{tot}$, where $P_{out}$ is the power transmitted to the even mode in the output NW and $P_{tot}$ is the total power emitted by the dipole.
The resulting electric field distribution is shown in Fig.~\ref{fig:dipole}(a), where emitter shows enhanced coupling to the output waveguide.

The optimal taper design spans four lattice sites in the $x$-direction (c.f. Fig.~\ref{fig:full_structure}(a)). The $y$-coordinates of each column of holes in the taper sections are linearly interpolated from the adjacent PCW sections, and three holes with decreasing radii of 60 nm, 55 nm, 50 nm along the $x$-direction are implemented at the center of Taper 1. The collection efficiency for different wavelengths are given in Fig.~\ref{fig:dipole}(c). At 934 nm, the dipole emission is greatly enhanced, as shown from the field distributions in Fig.~\ref{fig:dipole}(a), and the overall collection efficiency is $T_{col} = 0.89$. The device can be operated over a broad wavelength range, with collection efficiency $>0.85$ over a $10$-nm range. The slight increase of $T_{col}$ at short wavelengths is likely due to a better mode matching for the PCW-NW interface as $n_g$ becomes smaller for W1 PCW.

Additionally, we investigate the source efficiency as a function of the emitter offset along the $y$-direction ($\Delta y$), since a perfectly centered emitter will not couple to the excitation laser.
In the example shown in Fig.~\ref{fig:dipole}(b), i.e. when $\Delta y=70$ nm, the emitter couples to the odd mode of the input waveguide and the even mode of the output waveguide, as expected. Similar to $T_{col}$, the excitation efficiency, describing the coupling efficiency of the pump laser to the emitter, is defined as $T_{exc} = P_{in}/P_{tot}$, where $P_{in}$ is the power transmitted to the odd mode in the input NW. The collection and excitation efficiency can be calculated as a function of the offset, as shown in Fig.~\ref{fig:dipole}(d). As the emitter offset increases from $\Delta y=0$ to $70$ nm, the collection efficiency drops slightly to 0.82 and the excitation efficiency grows from 0 to 0.046. The relatively small excitation efficiency results mainly from the weak light-matter interaction with the laser mode as required for a near-unity collection efficiency, and also partly from a non-optimal interface between the NW and the mode-filter PCW, i.e., $T_{2in} < 1$, which can be further improved by adding a mode adapter section \cite{Rosa2005jlt}.

\begin{figure}
    \centering
    \includegraphics[width=13cm]{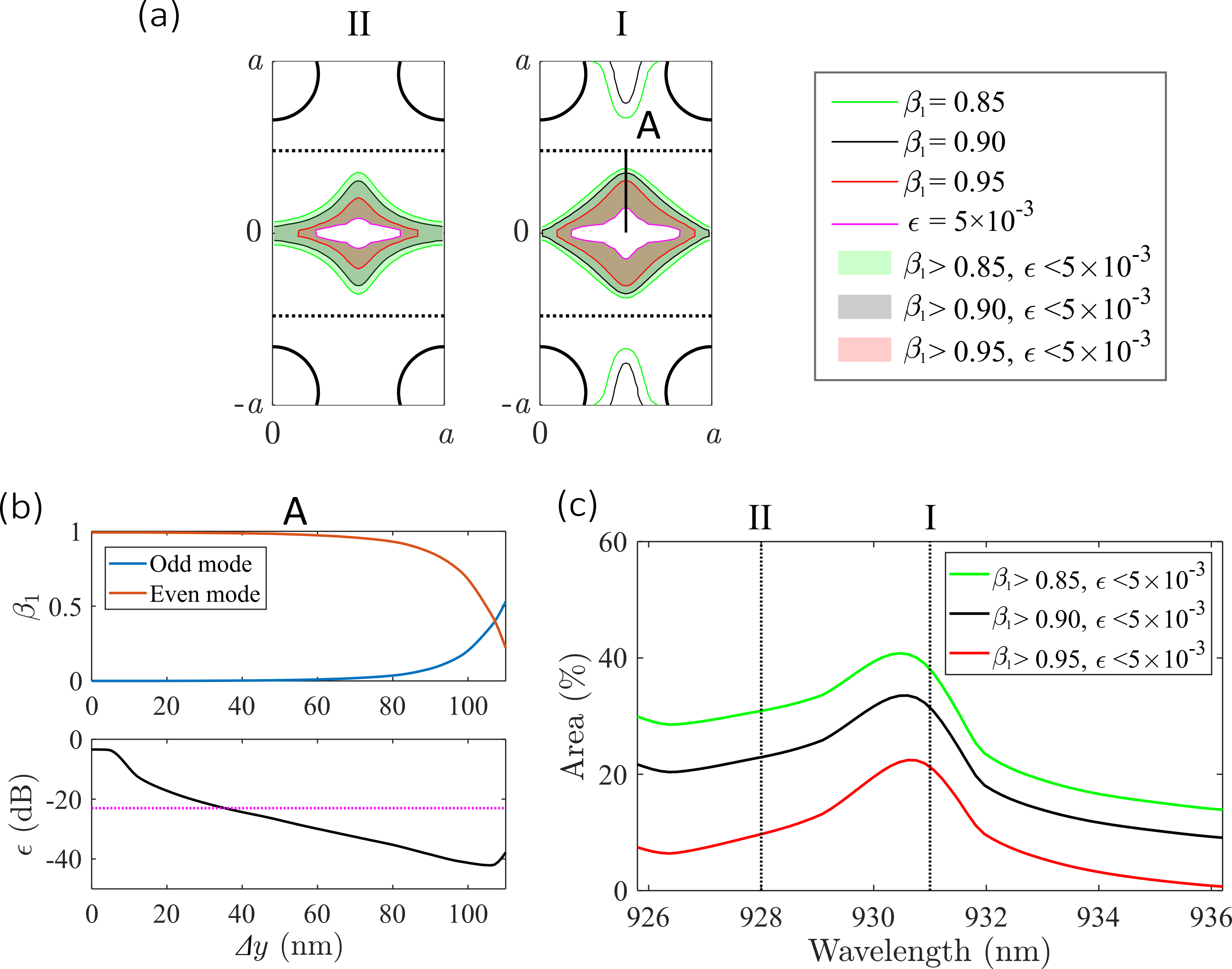}
    \caption{High purity (or low impurity $\epsilon$) and high collection efficiency ($\beta_1$) for single photons can be achieved simultaneously. (a) Isolines of $\epsilon = 5\times10^{-3}$ (magenta line) and $\beta_1 = 0.85$ (green line), 0.90 (black line), or 0.95 (red line) plotted together. The desired working areas are the shaded regions where $\epsilon < 5\times10^{-3}$ and $\beta_1 >$ 0.85 (light green), 0.90 (light black), and 0.95 (light red), respectively. The two panels labelled as I and II are for 931 nm and 928 nm (same as in Fig.~\ref{fig:maps}). (b) $\beta_1$ and $\epsilon$ for emitters as a function of its displacement in $y$-direction on the cutline A (see (a)) at $\lambda_{QD} = $ 931 nm. The magenta dotted line shows $\epsilon = 5\times10^{-3}$. (c) Area fraction of desired working region (calculated in a similar way as in Fig.~\ref{fig:maps}(d)) as a function of wavelength. }
    \label{fig:impurity}
\end{figure}

We conclude by discussing the conditions that allow achieving simultaneously a large $\beta$-factor and a low impurity $\epsilon$. $\epsilon$ is related to the second-order correlation function $g^{(2)}(t)$, which is measured experimentally via a Hanbury-Brown and Twiss setup, by $g^{(2)}(0) = 2\epsilon-\epsilon^2$ \cite{Kako2006natmat}. Here we consider $\epsilon = 5\times10^{-3}$ as a threshold value, which corresponds to a $g^{(2)}(0)$ of less than 0.01. Figure \ref{fig:impurity}(a) show areas where the impurity  $\epsilon > 5\times10^{-3}$, together with the regions where $\beta_1 > \beta_0$ (with $\beta_0 = $ 0.85, 0.90, or 0.95) for the two wavelength labeled as I and II shown in Fig.~\ref{fig:maps}(a). The calculation is performed in the case where the extinction $\eta = -50$ dB, as described above. In the magenta-line enclosed region, the emitter requires a too high excitation power to achieve the desired level of impurity. Thus, the desired working areas are the shaded regions with low impurity and high $\beta_1$. In Fig.~\ref{fig:impurity}(b), a cross-section of Fig.~\ref{fig:impurity}(a) is shown, where $\beta_1$, $\beta_2$, and the impurity $\epsilon$ are plotted as a function of the lateral offset of the QD. While $\epsilon$ drops below threshold of $5\times10^{-3}$, when $\Delta y>40$ nm, $\beta_1$ remains close to unity for a large range of $\Delta y$, implying that the areal fraction of the unit cell that simultaneously satisfies high $\beta$-factor and low-impurity is still large. This fraction is plotted in Fig.~\ref{fig:impurity}(c) as a function of wavelength. The results show that the device can operate over a broad range of QD wavelengths and distribution, with $\beta_1$ exceeding 0.95  and $g^{(2)}(0)<0.01$ for $\sim 22$\% of the effective unit cell area. Considering a typical QD density of 10 $\mu$m$^{-2}$, this means approximately 1 QD every 5 unit cells. Taking into account the inhomogeneous broadening of the emitters (standard deviation of around 10 nm, centered ad 930 nm), the density in the 3-nm range between wavelengths I and II (cf. Fig.~\ref{fig:impurity}(c)) reduces to roughly 1 $\mu$m$^{-2}$ and thus requires approximately 50 unit cells, or a total length of 12 $\mu$m to statistically find a QD. We believe that our new and robust single-photon source design will be a major step forward improving device yield, which is essential for ultimately scaling up to many QDs. To this end, also pre-localization of QDs followed by deterministic device fabrication is an important step forward \cite{pregnolato2020aplphoton}.

\section{Conclusion}\label{sec:conclusion}
We have demonstrated a novel cascaded PCW structure to realize efficient and stable excitation of QDs for single-photon sources, and conducted thorough numerical optimization of its performance. With band-structure engineering, a dual-mode PCW section has been designed for achieving almost unity $\beta$-factor ($\beta>0.95$) over large spectral and spatial ranges. Input and output PCW sections are designed and optimized, and an overall high laser extinction with single-photon impurity $\epsilon<5\times10^{-3}$ can be reached together with the high $\beta$-factor. Simulations show that overall single-photon collection efficiency $>0.89$ to the fundamental mode of the NW can be achieved in the designed device. The potentially stable excitation scheme, together with the very compact device footprint ($\sim 50$ $\mu$m$^2$), is highly desirable for quantum applications requiring, e.g., parallel single-photon sources and long strings of single photons from a single emitter.

\section{Acknowledgement}
 The authors gratefully acknowledge financial support from Danmarks Grundforskningsfond (DNRF 139, Hy-Q Center for Hybrid Quantum Networks), the National Natural Science Foundation of China (No. 62005195), Innovationsfonden (No. 9090-00031B, FIRE-Q), and the European Research Council (ERC) under the European Union’s Horizon 2020 research and innovation programme (Grant agreement No. 949043, NANOMEQ).

\bibliographystyle{unsrt}
\bibliography{references}

\end{document}